\documentclass[pre,aps,twocolumn,showpacs,superscriptaddress,floatfix]{revtex4-2}

\usepackage{mathtools}
\usepackage[usenames,dvipsnames]{xcolor}
\usepackage{amsmath}
\usepackage{amssymb,mathrsfs}
\usepackage{graphicx}
\usepackage[colorlinks=true]{hyperref}
\usepackage{soul}

\usepackage{float}

\newcommand{\prt}{\partial}

\newcommand{\eps}{\varepsilon}

\newcommand{\om}{\omega}

\newcommand{\sn}{\mathrm{sn}}

\begin{document}

\title{Evolution of localized pulses in the defocusing modified Korteweg--de Vries equation theory}

\author{L. F. Calazans de Brito}
\affiliation{Instituto de F\'{i}sica, Universidade de S\~{a}o Paulo,
05508-090 S\~{a}o Paulo, SP, Brazil}

\author{A. Gammal}
\affiliation{Instituto de F\'{i}sica, Universidade de S\~{a}o Paulo,
05508-090 S\~{a}o Paulo, SP, Brazil}
	
\author{A. M. Kamchatnov}
\affiliation{Institute of Spectroscopy, Russian Academy of Sciences, Troitsk, Moscow, 108840, Russia}
\affiliation{ Higher School of Economics, Physical Department, 20 Myasnitskaya ulica, Moscow, 101000, Russia}

\begin{abstract}
In this work, we develop, in the Gurevich-Pitaevskii framework, an analytic theory for the evolution of localized pulses in the defocusing modified Korteweg--de Vries equation theory for situations when a
dispersive shock does not eventually transform into a sequence of well-separated solitons. We found solutions to the Whitham modulation equations for the corresponding so-called ``quasi-simple'' dispersive
shock waves and illustrated this solution with concrete examples of an initial pulse. Comparison of the analytical solution with direct numerical simulations showed that the modulation theory provides a very
accurate description of the wave pattern even at one wavelength scale.
\end{abstract}



\maketitle

\section{Introduction}

The study of nonlinear phenomena has reached an advanced theoretical stage that explains a wide range of real-world physical phenomena on the basis of nonlinear wave models. Such advances can be seen, for example, in the development of the theory of {\it dispersive shock waves}~(DSWs) (see, e.g., review articles~\cite{elho06,kamchatnov21}), which consist of coherent nonlinear wave structures whose amplitudes change slowly from their small amplitude edges to soliton edges. These wave patterns form after the pulses' breaking in various media with competing dispersion and nonlinearity effects, so they were observed in such systems as plasma~\cite{plasma1,plasma2}, electron beams~\cite{electronBeams1}, optics~\cite{OpticalDSW1,OpticalDSW2},
Bose-Einstein condensates~\cite{DSW-BEC1,DSW-BEC2}, and others.

The theory of dispersive shocks was first developed by Gurevich-Pitaevskii~\cite{gurevich73}. In their work, they observed that the different scales between fast oscillations and slowly changing parameters along the DSW zone make the Whitham modulation theory~\cite{whitham65, whitham11} an appropriate mathematical tool for the analytical description of DSWs. As a result, one can find the slowly changing parameters as solutions of
first-order partial differential equations of hydrodynamic type (i.e., of the Whitham modulation equations). If the nonlinear wave equation for the model
under consideration is completely integrable, as it occurs in the Korteweg--de Vries~(KdV) equation case, then the Whitham equations can be cast into a diagonal form for the so-called Riemann invariants. In the simplest scenario with only one Riemann invariant changing along a DSW, we obtain a ``simple wave'' solution of the Whitham equations. This kind of structures can be observed, for example, in the so-called Riemann problem, where one studies the evolution of an initial step-like discontinuity.
This problem was considered for many completely integrable equations, in particular, for the KdV equation~\cite{gurevich73}, the modified Korteweg-de Vries~(mKdV) equation~\cite{el17}, the Gardner equation~\cite{kamchatnov12}, and the nonlinear Schr\"{o}dinger (NLS) equation~\cite{el95}. It should be
noticed that even in this case of simple-wave solutions, the modulation theory for the Gardner and mKdV equations with cubic nonlinearity turns out to be quite complicated in comparison with the KdV equation case.
Since these equations are not genuinely nonlinear~\cite{lax1957hyperbolic}, it turns out that one set of Riemann invariants, obtained as solutions of the Whitham equations, corresponds to two different wave patterns.

Although the simple-wave problems have many applications to situations that include the evolution of step-like discontinuities, this particular case is not applicable to a localized initial pulse when several Riemann invariants are changing along a DSW. Just this situation is considered, for instance,
in typical experiments on shallow water DSWs (see, e.g., Refs.~\cite{hammack74, trillo16} and references therein). The corresponding theory was developed in Refs.~\cite{gurevich89,el93,el02,isokam19} with two
Riemann invariants changing along DSWs for the KdV equation case, and such waves were called quasi-simple DSWs.
In this work, we develop the analytical theory of quasi-simple DSWs for the defocusing mKdV equation,
\begin{equation} \label{eq1}
	u_t -6u^2u_x + u_{xxx} = 0,
\end{equation}
in case of non-solitonic initial pulses and compare our analytical solution with a numerical experiment.

This article is organized as follows. Section~II is devoted to the calculation of the pulse evolution before the wave-breaking moment, i.e., the so-called gradient catastrophe. In section~III, we develop the Whitham theory and show how the evolution of Riemann invariants reproduces the shape of DSWs in the mKdV equation case. In section~IV, we study the motion of the DSW edges by the classical hodograph transform method. In section~V, we obtain parameters inside the DSW zone by the generalized Tsarev's hodograph method, and, finally, we solve some specific problems in section~VI in order to compare our analytical results with the direct numerical simulation.

\section{Non-dispersive limit for smooth pulse evolution}

\begin{figure}[t]
    \centering
    \includegraphics[width=8cm]{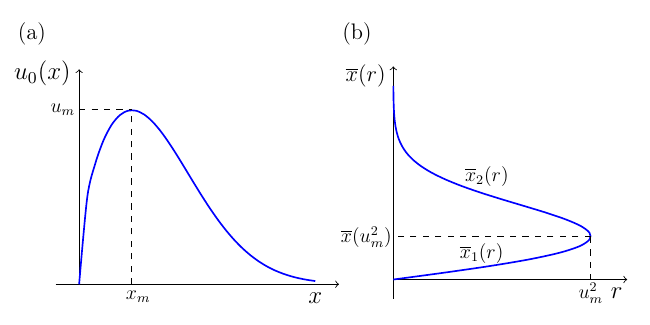}
    \caption{(a) Profile of a general quasi-simple pulse $u_0(x)$. (b) The inverse function of this initial
    condition defines two branches $\bar{x}_1(r)$ and $\bar{x}_2(r)$,
    where $r=u^2$. }
    \label{fig1}
\end{figure}

To solve the initial problem for the defocusing mKdV equation~\eqref{eq1}, we consider an arbitrary initial smooth pulse $u_0(x)$ with a single extremum value $u_m$ at the coordinate position $x_m$ (as shown in Fig.~\ref{fig1}(a)).
At the stage of evolution before the wave-breaking moment, the pulse deforms without loss of smoothness. In this regime, if the pulse with the amplitude $u_m$ has a long enough characteristic length ${l} \gg 1$, the third-order derivative term in Eq.~\eqref{eq1}, $u_{xxx} \sim u_m/{l}^3$, is small compared with the nonlinear term $u^2u_x\sim u_m^3/{l}$, and we can neglect it.
This means that the dispersive effects are much smaller than the nonlinear ones, and as a result, at this stage of evolution, the pulse evolves according to the Hopf equation
\begin{equation} \label{eq2}
    r_t - 6rr_x = 0,
\end{equation}
where we defined the variable $r = u^2$ for convenience of comparison with limiting values of the Riemann invariants in the theory of DSWs developed below.

Eq.~\eqref{eq2} can be solved by the standard method of characteristics, and the solution in implicit form reads
\begin{equation}\label{eq3}
	x + 6r t = \bar{x}_1(r) \quad \text{and} \quad x + 6r t = \bar{x}_2(r)
\end{equation}
where $\bar{x}_1(r)$ and $\bar{x}_2(r)$ represent the two branches of the inverse function of the initial distribution $r_0(x)=u_0^2(x)$, as depicted in Fig.~\ref{fig1}(b). These two branches evolve independently as long as the solution remains single-valued. However, the characteristics start to intersect each other at the instant of wave breaking,

\begin{equation} \label{eq4}
    t_{\text{WB}} = \frac{1}{6\max\left( dr / dx \right)},
\end{equation}

so the solution becomes multi-valued. Close to this moment, the dependence $u(x,t)$ on $x$ becomes very steep near the wave-breaking location, so the derivative becomes very large and the dispersive term of Eq.~\eqref{eq1}
cannot be neglected anymore. As a result, the interplay of dispersion and nonlinearity effects leads to the formation of a DSW. In the Gurevich-Pitaevskii approximation, the dispersive shock occupies a finite interval between the small-amplitude edge~($x_l$) and the soliton edge~($x_r$). As will be shown, this region expands
with time, and it matches with the dispersionless solution~(\ref{eq3}) at its edges. Within the DSW region, $x \in [x_l, x_r]$, the solution is obtained using the Whitham modulation theory, and outside those boundaries the shape of the pulse still evolves according to the solution (\ref{eq3}) of Eq.~\eqref{eq2}.

\section{Whitham modulation theory and DSW structure in mKdV equation}

The Whitham equations for Eq.~\eqref{eq1} were first found in Ref.~\cite{driscoll76} (see also Ref.~\cite{calkam24} for details) by averaging the conservation laws of Eq.~\eqref{eq1} over fast oscillations, followed by diagonalization of the averaged system. As a result, the modulation equations for the smooth functions $r_i(x, t)$ for $i\in\{ 1,2,3 \}$ are obtained, which change slowly along a DSW region. These variables $r_i(x, t)$ are called Riemann invariants of the Whitham system of modulation equations. In the case of quasi-simple solutions one of them is constant. To be definite, we assume that the Riemann invariants are ordered according to the inequalities
\begin{equation} \label{eq5}
    0=r_1 < r_2 \leq r_3.
\end{equation}
By taking $r_1$ equal to zero, we do not change the main characteristic features of DSWs, but simplify notation and calculations. Depending on the initial conditions, we obtain different wave patterns (see Refs.~\cite{calkam24,kamchatnov12,el17}), and we are interested in situations with the formation of non-solitonic pulses similar to what was found in Refs.~\cite{gurevich89, el93, el02, isokam19} for the KdV equation case. When the solution of the Whitham equations is found, the physical variables can be expressed in terms of them by the formulas
\begin{equation} \label{eq6}
        u(x,t) = \pm \frac{\sqrt{r_3} (\sqrt{r_3} - \sqrt{r_2}) - (r_3 - r_2)\sn^2(\theta/2, m)}{\sqrt{r_3} - (\sqrt{r_3} - \sqrt{r_2})\sn^2(\theta/2, m)},
\end{equation}
where the modulus $m$ of elliptic functions and the wave phase $\theta$ are also functions of the Riemann invariants.
The parameter $m$ satisfies the condition $0 \leq m \leq 1$, and it is defined by the expression
\begin{equation} \label{eq7}
    m =1- \frac{r_2}{r_3},
\end{equation}
while the phase has the usual form
\begin{equation} \label{eq8}
    \theta (x,t) = kx - \om t +\theta_0,
\end{equation}
where the local wave number and wave frequency are given, respectively, by the expressions
\begin{equation}\label{eq9}
    k = 2 \sqrt{r_3 } \quad \text{and} \quad \om = kV,
\end{equation}
with the phase velocity
\begin{equation} \label{eq10}
    V = -2(r_2 +r_3)
\end{equation}
of the nonlinear periodic wave.

Since we take $r_1=0$ and only $r_2$ and $r_3$ change along a DSW, the Whitham system reduces to two equations
\begin{equation} \label{eq11}
\begin{aligned}
    \frac{\partial r_2}{\partial t} + v_2(r_2, r_3)\,\frac{\partial r_2}{\partial x} &= 0, \\
    \frac{\partial r_3}{\partial t} + v_3(r_2, r_3)\,\frac{\partial r_3}{\partial x} &= 0,
\end{aligned}
\end{equation}
where the Whitham velocities are defined as
\begin{equation} \label{eq12}
    v_i(r_2, r_3) = \left( 1 - \frac{L}{\prt L /\prt r_i} \frac{\prt}{\prt r_i}\right)V.
\end{equation}
Here, $L$ is the local wavelength of the nonlinear periodic solution (\ref{eq6}),
\begin{equation} \label{eq13}
     L = \frac{2}{\sqrt{r_3}}K(m),
\end{equation}
where $K(m)$ is the complete elliptic integral of the first kind.
Substituting Eq.~\eqref{eq13} into Eq.~\eqref{eq12}, we can write $v_i$ as
\begin{equation} \label{eq14}
\begin{aligned}
	&v_2(r_2, r_3) = -2(r_2+r_3) - \frac{4(1-m)(r_3 - r_2)K(m)}{E(m)-(1-m)K(m)}, \\
	&v_3(r_2, r_3) = -2(r_2+r_3) + \frac{4mr_3 K(m)}{E(m)-K(m)},
\end{aligned}
\end{equation}
where $E(m)$ denotes the complete elliptic integral of the second kind.
In contrast to the KdV equation case, where the evolutions of positive and negative pulses are qualitatively different, in the mKdV equation case the positive and negative localized pulses evolve symmetrically with respect to the $x$-axis. By definition, we have $r = u^2> 0$, so the distributions of the same set of Riemann invariants provide solutions for both positive and negative initial pulse.

\section{Hodograph transform and expansion of DSW region}

As mentioned above, the DSW region expands during its evolution. In this section, we will develop the theory of the DSW expansion by means of the hodograph transform.

\subsection{Hodograph transform}

The main idea behind the hodograph technique consists of ``swapping'' the independent and dependent variables in the system~\eqref{eq11}, that is
\begin{equation} \label{eq15}
    (r_2(x,t), r_3(x,t)) \rightarrow (x(r_2, r_3), t(r_2, r_3)).
\end{equation}
We assume that the Jacobian
\begin{equation} \label{eq16}
    J = \frac{\prt x}{\prt r_2}\frac{\prt t}{\prt r_3} - \frac{\prt x}{\prt r_3}\frac{\prt t}{\prt r_2}
\end{equation}
does not vanish, $J \neq 0$. Then, the transformation to new variables is performed by means of the expressions
\begin{equation} \label{eq17}
    \begin{aligned}
        &\frac{\prt r_2}{\prt x} = J^{-1}\frac{\prt t}{\prt r_3}, \quad \frac{\prt r_2}{\prt t} = - J^{-1}\frac{\prt x}{\prt r_3}, \\
         &\frac{\prt r_3}{\prt x} = -J^{-1}\frac{\prt t}{\prt r_2}, \quad \frac{\prt r_3}{\prt t} =  J^{-1}\frac{\prt x}{\prt r_2},
    \end{aligned}
\end{equation}
and after substitution of Eqs.~\eqref{eq17} into Eq.~\eqref{eq11}, we obtain the system
\begin{equation} \label{eq18}
\begin{aligned}
    \frac{\partial x}{\partial r_2} - v_3(r_2, r_3)\,\frac{\partial t}{\partial r_2} &= 0, \\
    \frac{\partial x}{\partial r_3} - v_2(r_2, r_3)\,\frac{\partial t}{\partial r_3} &= 0.
\end{aligned}
\end{equation}
It must be fulfilled along the whole dispersive shock, but at first, we will study it at the shock's edges in order to find their paths.

\subsection{Path of the soliton edge}

\begin{figure}[t]
    \centering
    \includegraphics[width=8cm]{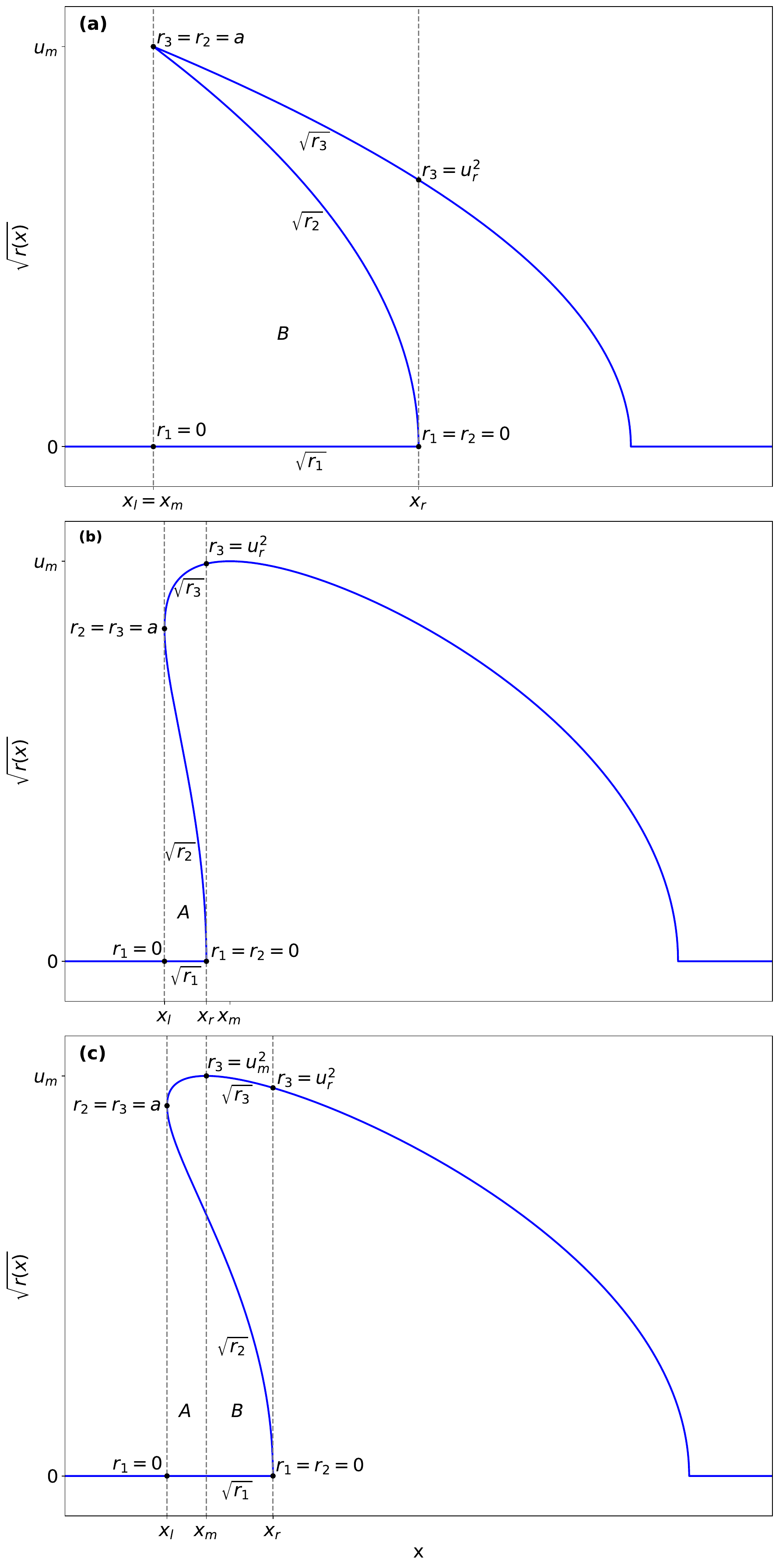}
    \caption{Sketch of the distribution of the square root of Riemann invariants for a DSW in the interval $[x_l, x_r]$. (a) When $a = u_m^2$ is constant throughout the evolution, then $x_l = x_m$ and the shock lies entirely within region $B$. Panels (b) and (c) represent a more general situation where $a < u_m^2$ at different stages: (b) before the soliton edge reaches the second branch $\bar{x}_2(r)$, the DSW is located in region $A$ to the left of $x_m$; (c) after it enters this branch, the shock requires two distinct analytical descriptions, one for region $A$ ($[x_l, x_m]$) and another for region $B$ ($[x_m, x_r]$).}
    \label{fig2}
\end{figure}

At the soliton edge, $x_r$, we have $r_2 \rightarrow r_1 = 0$ and $r_3 \rightarrow r = u_r^2$ (see Fig.~\ref{fig2}). Substituting these values in Eqs.~\eqref{eq14}, we find that at this boundary $v_2(0, r) = -2r$ and $v_3 (0, r) = -6r$. So, the first Eq.~\eqref{eq18} reduces to
\begin{equation} \label{eq19}
    \frac{\prt }{\prt r_2}(x + 6rt) = 0,
\end{equation}
from which we reproduce the background solution for the smooth limit shown in Eq.~\eqref{eq3},
\begin{equation} \label{eq20}
    x + 6rt = \bar{x}_i(r).
\end{equation}
At this edge, the second equation depends only on $r$, which is the value of the Riemann invariant at the location of the DSW's edge, so along the path of the edge we have $\prt/\prt r_3 \rightarrow d/dr$, and
\begin{equation}\label{eq21}
    \frac{d x}{d r} + 2r\frac{dt}{dr} = 0.
\end{equation}
Substituting the value of $x(r)$ from Eq.~\eqref{eq20} in Eq.~\eqref{eq21}, we obtain the differential equation
\begin{equation} \label{eq22}
    \frac{dt}{dr} +\frac{3t}{2r} = \frac{1}{4r}\frac{d}{dr}\bar{x}_i(r),
\end{equation}
which can be readily solved to give
\begin{equation}\label{eq23}
    t(r_r) = \left(\frac{r_0}{r_r}\right)^{3/2}t_{\text{WB}} +
    \frac{1}{4r_r^{3/2}}\int_{r_0}^{r_r} \sqrt{r}\frac{d }{dr} \bar{x}_i(r) dr,
\end{equation}
where $r_0$ represents the wave amplitude at the breaking point $t_{\text{WB}}$, and $r_r$ is the background value at the position of the soliton edge. The coordinate of the soliton edge is obtained after the substitution of Eq.~\eqref{eq23} into Eq.~\eqref{eq20},
\begin{equation} \label{eq24}
    x(r_r) = -6r_rt(r_r) + \bar{x}_i(r_r).
\end{equation}

\subsection{Path of the small-amplitude edge}

At the small amplitude edge, $x_l$, $r_1 \rightarrow u^2 = 0$, $u$ being here the vanishing amplitude of the background wave. At this edge, in the Gurevich-Pitaevskii approximation, the Riemann invariants $r_2$ and $r_3$ take the same value which we denote as $a$, $r_2 = r_3 = a$. Then we get from Eq.~\eqref{eq14}, that $v_2 = v_3 = -12a$, so both hodograph equations~\eqref{eq18} reduce to
\begin{equation} \label{eq25}
    \frac{d x}{da} + 12a\frac{dt}{da} = 0,
\end{equation}
that is, the small-amplitude edge propagates along the characteristic line defined by the equation
\begin{equation}\label{eq26}
    \frac{d x_l}{dt} = -12a.
\end{equation}
Actually, this is the group velocity of the packet associated with the small-amplitude edge. Indeed, the linearized Eq.~\eqref{eq1} gives the dispersion relation
\begin{equation} \label{eq27}
    \om = -6u^2k - k^3,
\end{equation}
so, at this edge, with zero background amplitude $u \rightarrow 0$, the group velocity is equal to
\begin{equation} \label{eq28}
    v_g = \frac{\prt \om}{\prt k} = -3k^2.
\end{equation}
From Eq.~\eqref{eq9} we find that here we have $k = 2\sqrt{a}$, so we get again Eq.~\eqref{eq26}.

Depending on the initial profile of the pulse, we get two different situations. If at the time of wave breaking
we have $a = u_m^2$ (see Fig.~\ref{fig2}(a)), then the value of $a$ remains constant during the
evolution. Consequently, Eq.~\eqref{eq26} yields the motion of the small-amplitude edge with constant velocity,
\begin{equation} \label{eq29}
    x_l(t) = -12u_m^2 (t - t_{\text{WB}}).
\end{equation}
The second more general possibility corresponds to $a < u_m^2$ (see Fig.~\ref{fig2}(b)-(c)), so $a$ changes during the evolution of the pulse. In this case, as shown in Fig.~\ref{fig2} (b) the DSW is first in the region on the left from the maximum $u_m$, that is, the soliton edge propagates along the first branch $\overline{x}_1(r)$ of the inverse function of the initial distribution; we say that this edge is located inside the region $A$. At some instant of time, the soliton edge reaches the maximum of the distribution $u=u_m$, and after that it propagates along the second branch $\overline{x}_2(r)$ of the inverse function of the initial distribution. Hence, the DSW consists of two parts: region $A$ on the left from the maximal point $u_m^2$ of two matching Riemann invariants $r_2$ and $r_3$, and region $B$ on the right from this maximal point of two matching Riemann invariants. This case is depicted in Fig.~\ref{fig2}(c). Actually, the first case (a) corresponds to situations when the region $A$ shrinks to a point, so there remains only the region $B$.
In case (b), we cannot find the function $a(t)$ just from the limiting
equation (\ref{eq25}), so we have to solve the hodograph equations along the whole DSW, and solution of this problem is necessary for the analytical description of the DSW's profile.

\section{General quasi-simple solution of Whitham Equations}

In this section, we use the method of Refs.~\cite{gurevich92,isokam19} to explicitly solve the Whitham equations~\eqref{eq11} in the region $\left(x_l, x_r \right)$ (see Fig.~\ref{fig2}). As a result, we find
the distributions of $r_2(x, t)$ and $r_3(x,t)$, which determine the modulated wave function~\eqref{eq6}.

\subsection{Generalized hodograph method and Euler-Poison-Darboux equation}

The generalized hodograph method, proposed by Tsarev in Ref.~\cite{tsarev91}, is based on the assumption that the solution of Eqs.~\eqref{eq11} can be found implicitly in the form
\begin{equation} \label{eq31}
    \begin{aligned}
        &x - v_2(r_2,r_3) t = w_2(r_2, r_3), \\
        &x - v_3(r_2,r_3) t = w_3(r_2, r_3).
    \end{aligned}
\end{equation}
These equations combined with Eqs.~\eqref{eq18} yield an important relationship
\begin{equation} \label{eq32}
    \frac{1}{w_i - w_j}\frac{\prt w_i}{\prt r_j} = \frac{1}{v_i - v_j}\frac{\prt v_i}{\prt r_j}.
\end{equation}
The symmetry of this relationship with respect to the transposition $v_i\leftrightarrow w_i$ suggests that we can find the solution of $w_i(r_2, r_3)$ in the form similar to $v_i(r_2, r_3)$ in Eq.~(\ref{eq12}), that is
\begin{equation} \label{eq33}
    w_i(r_2, r_3) = \left( 1 - \frac{L}{\prt L /\prt r_i} \frac{\prt}{\prt r_i}\right) W(r_2, r_3).
\end{equation}
Then, following the methods of Refs.~\cite{gurevich92,kamchatnov21,kamchbook}, we find from Eqs.~\eqref{eq32},
\eqref{eq33}, and \eqref{eq12}, that the function $W(r_2, r_3)$ must satisfy the Euler-Poisson-Darboux equation
\begin{equation} \label{eq34}
    \frac{\prt^2 W}{\prt r_2 \prt r_3} = \frac{1}{2(r_3 - r_2)}\left( \frac{\prt W}{\prt r_3} - \frac{\prt W}{\prt r_2} \right).
\end{equation}
Consequently (see Refs.~\cite{gurevich92,kamchatnov21,kamchbook}), the general solution of Eq.~\eqref{eq34} can be
written in the form
\begin{equation} \label{eq35}
    \begin{aligned}
        W(r_2, r_3) = &\int_0^{r_2} \frac{\Phi(\mu) d\mu}{\sqrt{(r_3 - \mu)(r_2-\mu)}} \\
        &+ \int_0^{r_3} \frac{\Psi(\mu) d\mu}{\sqrt{(r_3 - \mu)|r_2-\mu|}},
    \end{aligned}
\end{equation}
where $\Phi(\mu)$ and $\Psi(\mu)$ are arbitrary functions to be found from the appropriate boundary conditions.
Therefore, the solution depends on the position of the soliton edge. If the soliton edge is located in the branch $A$, we denote it as $W^A(r_2, r_3)$, otherwise, if the soliton edge is moving through the second branch,
the solution is denoted as $W^B(r_2, r_3)$.

Immediately after the wave-breaking moment, the soliton edge starts its motion along the  branch $A$.
At this boundary, we have $r_2 \rightarrow 0$, so Eq.~(\ref{eq35}) becomes
\begin{equation} \label{eq36}
    W^A(0, r_3) = \int_0^{r_3}\frac{\Psi^A(\mu) d\mu}{\sqrt{\mu(r_3 - \mu)}}.
\end{equation}
This is well-known Abel equation, so its solution is given by the formula
\begin{equation} \label{eq37}
    \frac{\Psi^A(\mu)}{\sqrt{\mu}} = \frac{1}{\pi}\frac{d}{d\mu}\int_0^\mu \frac{W^A(0, r)}{\sqrt{\mu - r}}dr.
\end{equation}
The function $W^A(0,r)$ can be found from the second equation of the system \eqref{eq31}.
In the soliton limit this equation reduces to
\begin{equation} \label{eq38}
    x + 6rt = w^A_3(0,r),
\end{equation}
so, while the soliton edge is located in the region $A$, that is $x=\bar{x}_1(r)$, this equation coincides with Eq.~\eqref{eq20}. Consequently, we get
\begin{equation} \label{eq39}
    w^A_3(0,r) = \bar{x}_1(r),
\end{equation}
and then Eq.~\eqref{eq33} converts in this limit into the differential equation
\begin{equation} \label{eq40}
    \bar{x}_1(r) =  \left( 1+ 2r\frac{\prt}{\prt r} \right) W^A(0,r).
\end{equation}
This equation can be readily solved,
\begin{equation} \label{eq41}
    W^A(0,r) = \frac{1}{2\sqrt{r}}\int_0^r\frac{\bar{x}_1(\rho)}{\sqrt{\rho}}d\rho.
\end{equation}
so we can rewrite Eq.~\eqref{eq37} as
\begin{equation} \label{eq42}
    \frac{\Psi^A(\mu)}{\sqrt{\mu}} = \frac{1}{2\pi}\frac{d}{d\mu}\int_0^\mu
    \left( \frac{1}{\sqrt{r(\mu - r)}} \int_0^r\frac{\bar{x}_1(\rho)}{\sqrt{\rho}} d\rho\right) dr.
\end{equation}
We change the order of integration,
\begin{equation} \label{eq43}
    \frac{\Psi^A(\mu)}{\sqrt{\mu}} = \frac{1}{2\pi}\frac{d}{d\mu}\int_0^\mu \left(
    \frac{\bar{x}_1(\rho)}{\sqrt{\rho}}\int_\rho^\mu  \frac{dr}{\sqrt{r(\mu - r)}} \right) d\rho,
\end{equation}
where the inner integral is elementary,
\begin{equation} \label{eq44}
    \int_\rho^\mu  \frac{dr}{\sqrt{r(\mu - r)}} = \arccos\left(\frac{2\rho}{\mu} - 1\right).
\end{equation}
Thus, we arrive at the expression
\begin{equation} \label{eq45}
    \frac{\Psi^A(\mu)}{\sqrt{\mu}} = \frac{1}{2\pi}\frac{d}{d\mu}\int_0^\mu
    \frac{\bar{x}_1(\rho)}{\sqrt{\rho}} \arccos\left(\frac{2\rho}{\mu} - 1\right) d\rho,
\end{equation}
or
\begin{equation}\label{eq46}
    \Psi^A(\mu) = \frac{1}{2\pi\sqrt{\mu}}\int_0^\mu \frac{\bar{x}_1(\rho)}{\sqrt{\mu - \rho}} d\rho.
\end{equation}
It is easy to show that if  $\Psi(r) + \Phi(r) \neq 0$, then Eq.~(\ref{eq35}) with $r_3=r_2+\varepsilon$ diverges in the limit $\varepsilon\to0$, so we must have $\Psi(r)=-\Phi(r)$, and we arrive at the final expression for the solution of the Euler-Poisson-Darboux equation which satisfies the boundary conditions corresponding to the DSW located in the region $A$:
\begin{equation}\label{eq47}
    W^A(r_2, r_3) = \int_{r_2}^{r_3} \frac{\Psi^A(\mu) d\mu}{\sqrt{(r_3 - \mu)(\mu - r_2)}}.
\end{equation}

The soliton edge enters the region $B$ at the point with $r_3=u_m^2$, and after that moment the DSW consists of two parts. In the region $A$ the solution is still given by the formulas (\ref{eq31}), (\ref{eq33}) with $W=W^A(r_2,r_3)$ defined by Eq.~(\ref{eq47}). The solution in the region $B$ corresponds to another function $W=W^B(r_2,r_3)$, which must satisfy the boundary condition $W^A(r_2, u_m^2)= W^B(r_2, u_m^2)$. Therefore, we look for the solution of the Euler-Poisson-Darboux equation in the form
\begin{equation}\label{eq48}
    W^B(r_2, r_3) = W^A(r_2, r_3)  +\int_{r_3}^{u_m^2} \frac{\Psi^B(\mu) d\mu}{\sqrt{(\mu-r_2)(\mu - r_3)}}.
\end{equation}
Then at the soliton edge ($r_2 \rightarrow 0$) we get
\begin{equation} \label{eq49}
    W^B(0, r_3) - W^A(0, r_3)  =\int_{r_3}^{u_m^2} \frac{\Psi^B(\mu) d\mu}{\sqrt{\mu(\mu-r_3) }}.
\end{equation}
This is the Abel equation again, so the inverse Abel's transform yields
\begin{equation}\label{ak-1}
  \frac{\Psi^B(\mu)}{\sqrt{\mu}} = -\frac{1}{\pi}\frac{d}{d\mu}\int_\mu^{u_m^2} \frac{W^B(0, r)-W^A(0,r)}{\sqrt{r - \mu}}dr.
\end{equation}
Then, for the propagation of the soliton edge along the second branch, instead of Eq.~(\ref{eq40}), we get
\begin{equation} \label{ak-2}
    \bar{x}_2(r) =  \left( 1+ 2r\frac{\prt}{\prt r} \right) W^B(0,r)
\end{equation}
which must be solved with the boundary condition $W^B(0,u_m^2)=W^A(0,u_m^2)$. Thus, taking into account Eq.~(\ref{eq41}), we obtain
\begin{equation} \label{ak-3}
    W^B(0,r) = -\frac{1}{2\sqrt{r}}\int_r^{u_m^2}\frac{\bar{x}_2(\rho)}{\sqrt{\rho}}d\rho+
    \frac{1}{2\sqrt{r}}\int_0^{u_m^2}\frac{\bar{x}_1(\rho)}{\sqrt{\rho}}d\rho.
\end{equation}
We subtract Eq.~(\ref{eq41}) from this equation and arrive at
\begin{equation}\label{ak-4}
  W^B(0, r) - W^A(0, r)  =-\frac{1}{2\sqrt{r}}\int_r^{u_m^2}\frac{\bar{x}_2(\rho)-\bar{x}_1(\rho)}{\sqrt{\rho}}d\rho.
\end{equation}
By substituting this expression into Eq.~(\ref{ak-1}), changing the order of integrations, and simplifying, we arrive at the final expression for the function $\Psi^B(\mu)$:
\begin{equation}\label{eq50}
    \Psi^B(\mu) = -\frac{1}{2\pi\sqrt{\mu}}\int_\mu^{u_m^2} \frac{\bar{x}_2(\rho) - \bar{x}_1(\rho)}{\sqrt{\rho - \mu }} d\rho.
\end{equation}
Subsequently, substituting Eq.~\eqref{eq50} into Eq.~\eqref{eq48} yields the solution to the Euler-Poisson-Darboux
equation for region $B$. The formulas derived in this section provide the complete solution for the Whitham equations,
expressed in terms of the initial distribution $r=u_0^2(x)$ and the two branches of its inverse functions $\bar{x}_1(r)$ and $\bar{x}_2(r)$.

\subsection{Phase shift correction}

When the Riemann invariants $r_2=r_2(x,t), r_3=r_3(x,t)$, as slow functions of $x$ and $t$, are found, we can easily
calculate the other envelope functions of the DSW, for example its amplitude. In particular, the wave number $k$ and
the frequency $\om=kV$ in the phase (\ref{eq8}) are given by Eqs.~(\ref{eq9}), (\ref{eq10}). However, the phase
shift $\theta_0$ in Eq.~(\ref{eq8}) also becomes a slow function of $x$ and $t$ in a modulated DSW. It turns out
that this function can also be expressed in terms of the same Riemann invariants $r_2$ and $r_3$ (see, e.g., Ref.~\cite{elho06,kpt08,huang24}).
To do this, we differentiate Eq.~(\ref{eq8}) with respect to $x$ to obtain
\begin{equation} \label{eq52}
    \frac{\prt \theta}{\prt x} = k + \sum_{i=2}^3 \left[ \frac{\prt k}{\prt r_i} x -
    \frac{\prt \omega}{\prt r_i} t + \frac{\prt \theta_0}{\prt r_i} \right] \frac{\prt r_i}{\prt x}
\end{equation}
By definition, we have $\prt \theta/\prt x = k$, and since the Riemann invariants $r_2,r_3$ are independent of each other, we get
\begin{equation} \label{eq53}
     \frac{\prt k}{\prt r_i} x - \frac{\prt \omega}{\prt r_i} t + \frac{\prt \theta_0}{\prt r_i} = 0.
\end{equation}
Using Eq.~(\ref{eq12}) we find
\begin{equation}\label{ak-5}
  \frac{\prt\om / \prt r_i}{\prt k/\prt r_i}=\frac{\prt (kV) / \prt r_i}{\prt k / \prt r_i}=\left(1+\frac{k}{\prt k/\prt r_i}\frac{\prt}{\prt r_i}\right)V = v_i.
\end{equation}
Now we divide Eq.~(\ref{eq53}) by $\prt k / \partial r_i$, and substitute Eq.~\eqref{ak-5} in it to obtain
\begin{equation}\label{ak-6}
  x-v_it=-\frac{\prt \theta_0 / \prt r_i}{\prt k/ \prt r_i}.
\end{equation}
Comparison with the hodograph solution of the Whitham equations (see Eqs.~(\ref{eq31}), (\ref{eq33}))
gives
\begin{equation} \label{eq54}
    -\frac{\prt \theta_0}{\prt r_i} = \frac{\prt}{\prt r_i} \left( k  W \right),
\end{equation}
and, after evident integration, we find the explicit expression for the phase correction
\begin{equation} \label{eq55}
    \theta_0 = - k W + C.
\end{equation}
So, we can rewrite the expression \eqref{eq8} for the phase as
\begin{equation} \label{eq56}
    \theta(x, t) = kx - \omega t  - k  W + C,
\end{equation}
where the integration constant $C$ is the initial wave shift parameter which is defined by the wave-breaking coordinate, and $W$ is either $W^A$ or $W^B$, depending on the spatial region where the solution is being evaluated.

\section{Analytical and Numerical Solutions: Comparison and Discussion}
\label{comparison}

In this section, we compare our analytical theory with a direct numerical experiment. In our numerical simulations, we use the fourth-order split-step Fourier method for the mKdV equation, similar to the approach of Refs.~\cite{muslu03, deiterding13}. To avoid artifacts caused by  periodic boundary conditions, we employ a large computational domain, $x \in [-1500, 1500]$, with $N=2^{16}$ equally spaced points; the time step used is $\Delta t = 0.001$.

We apply this method to the description of a quasi-simple DSW evolved from the initial condition
\begin{equation}  \label{eq57}
     u(x,0) =
    \begin{cases}
        \pm \sqrt{\frac{x}{60}}, \qquad 0 \leq x \leq 60, \\
        \pm\sqrt{\frac{120 - x}{60}}, \qquad 60 \leq x \leq 120, \\
        0, \qquad \text{otherwise},
    \end{cases}
\end{equation}
where the sign represents a ``positive'' and ``negative'' initial pulse, respectively. Actually, the transformation $u^2(x,t) \rightarrow r(x,t)$ shows that both positive and negative initial pulses are described by the same solution in terms of the Riemann invariants. This choice of the initial distribution simplifies calculations and it is convenient for comparing with the analytical theory since, in this case, the wave breaking leads to the quasi-simple DSWs with the pattern of the Riemann invariants shown in Fig.~\ref{fig2}(a). Consideration of a more involved general case corresponding to Figs.~\ref{fig2}(b,c) is relegated to Appendix~\ref{app:general_case}.

\begin{figure*}[t]
    \centering
    \includegraphics[width=17cm]{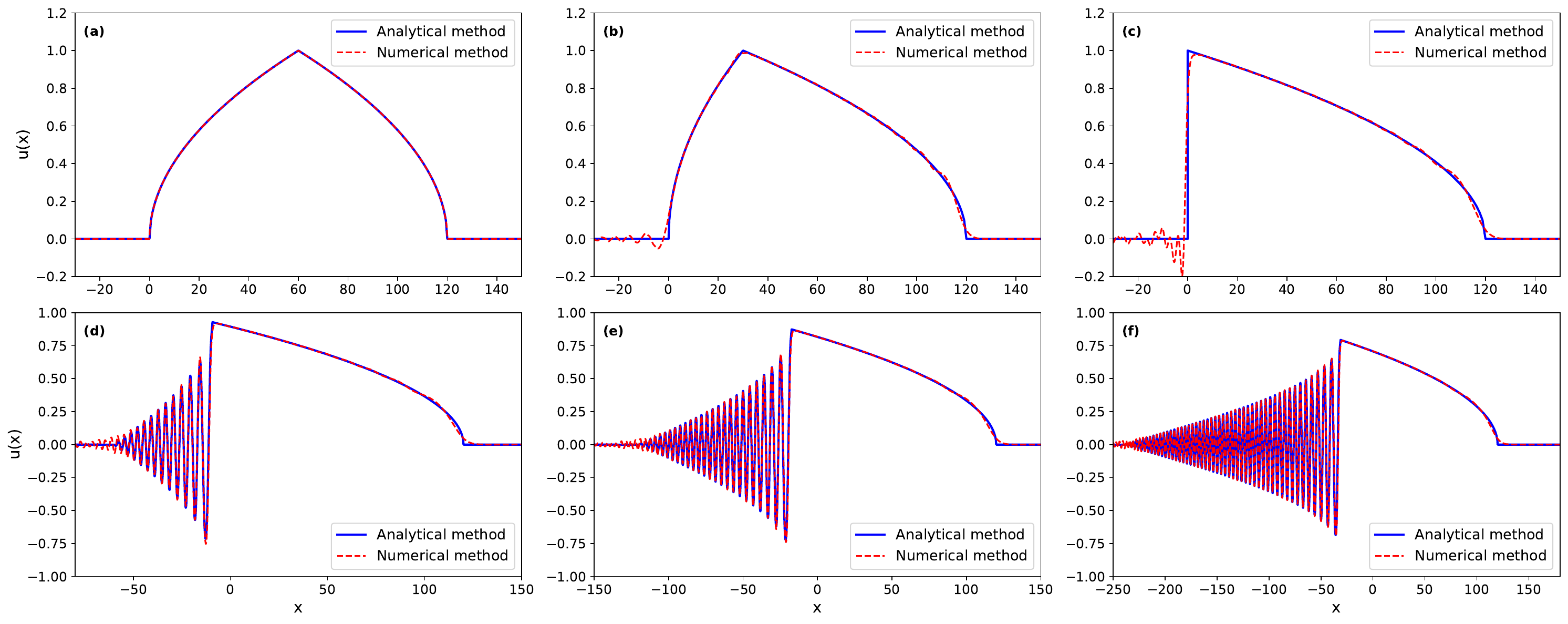}
    \caption{Evolution of a quasi-simple positive pulse for the mKdV equation at times (a) $t=0$, (b) $t=5$, (c) $t=10$, (d) $t=15$, (e) $t=20$, and (f) $t=25$. The plots show a comparison between numerical results obtained via the split-step Fourier method (dashed red lines) and analytical solutions (solid blue lines) derived from Eq.~\eqref{eq6} using the Riemann invariants.}
    \label{fig3}
\end{figure*}

\subsection{Non-dispersive regime and wave breaking}

Approximately up to the wave breaking moment, the analytical and numerical solutions for the evolution of a positive pulse in Eq.~\eqref{eq57} are shown in Fig.~\ref{fig3}(a)-(c). In this regime, the analytical solution of each branch satisfies the Hopf equation~\eqref{eq2} with the initial conditions described by the two branches
\begin{equation} \label{eq58}
    \bar{x}_1(r) = 60r, \quad \bar{x}_2(r) = 120 - 60r
\end{equation}
of the inverse function of Eq.~\eqref{eq57}, where $r = u^2$. As a result, from Eq.~\eqref{eq3}, we find that the solution of the Hopf equation reads
\begin{equation} \label{eq59}
    x_1(r, t) = 6r(10 - t), \quad   x_2(r, t) = 120 - 6r(t + 10).
\end{equation}
It is valid until the breaking time moment $t<t_{\text{WB}}$ of the first branch of Eq.~\eqref{eq57}, where
$dr/d\bar{x}_1 = 1/60$, so we find from Eq.~\eqref{eq4} that
\begin{equation} \label{eq60}
    t_{\text{WB}} = 10.
\end{equation}
As one can see in Fig.~\ref{fig3}(c), in the vicinity of the wave-breaking coordinate, a DSW starts its evolution, although the Hopf equation describes the wave profile quite well almost everywhere far enough from the wave-breaking location. After a long enough time of evolution, a DSW becomes long, and, according to the Gurevich and Pitaevskii approach, the Whitham modulation theory can be used for its analytical description.

\subsection{Solution of the dispersive shock problem}

After the wave-breaking moment, all parameters of the DSW can be found from the calculation of the Riemann invariants as solutions of the Whitham modulation equations. Here we find these Riemann invariants for the DSW evolving from the initial condition~\eqref{eq57}. As mentioned above, the dynamics of the DSW is treated differently depending on the region where the soliton edge of the DSW is located, namely, in the region $A$ or $B$. In the case of the region $A$, we find the potential function $W^A(r_2, r_3)$ from Eq.~\eqref{eq47}, where in our case Eq.~\eqref{eq46} gives the expression
\begin{equation} \label{eq61}
    \Psi^A(\mu) = \frac{1}{2\pi\sqrt{\mu}}\int_0^\mu\frac{60 \rho}{\sqrt{\mu-\rho}}d\rho=
    \frac{40\mu}{\pi}.
\end{equation}
Consequently, an easy calculation gives
\begin{equation} \label{eq63}
\begin{split}
    W^A(r_2, r_3)& = \frac{40}{\pi}\int_{r_2}^{r_3}\frac{\mu d\mu}{\sqrt{(r_3-\mu)(\mu-r_2)}}\\
   & = 20(r_2 + r_3),
    \end{split}
\end{equation}
so we get from Eq.~\eqref{eq33} that
\begin{equation} \label{eq65}
    w_i^A(r_2, r_3) = 20(r_2 + r_3) - 20 \left(  \frac{v_i}{2} + r_2 + r_3\right).
\end{equation}
This solution exists as long as the leading soliton does not reach the branch $B$, 
and this moment corresponds to the limit
\begin{equation} \label{eq67}
    r_2 \rightarrow 0,\quad r_3 \rightarrow r,\quad v_2 \rightarrow -2r,\quad  v_3 \rightarrow -6r.
\end{equation}
Hence, we get
\begin{equation} \label{eq68}
    w^A_2(0, r) = 20 r \quad \text{and} \quad w^A_3(0, r) = 60 r,
\end{equation}
so formally the soliton reaches the branch $B$ at the moment $t$ and at the point $x_r$ which satisfies the system
\begin{equation} \label{eq69}
        x_r + 2rt = 20r, \qquad
        x_r +6rt = 60r.
\end{equation}
Its solution obviously gives $t = 10 = t_{\text{WB}}$, that is, the soliton edge reaches the branch $\bar{x}_2(r)$ exactly at the moment of the wave breaking; see Fig.~\ref{fig3}(c). So, in our particular case with
the initial distribution (\ref{eq57}), the evolution of the DSW is described only by the parameters corresponding to the region $B$. In this region, we find from Eq.\eqref{eq50}
\begin{equation} \label{eq70}
\begin{split}
    \Psi^B(\mu)& = -\frac{60}{\pi\sqrt{\mu}}\int_\mu^1 \frac{1-\rho}{\sqrt{\rho - \mu}}d\rho\\
   & = -\frac{80}{\pi \sqrt{\mu}}(1-\mu)^{\frac{3}{2}}
    \end{split}
\end{equation}
Consequently, Eq.\eqref{eq48} yields
\begin{equation} \label{eq72}
    \begin{split}
        W^B(r_2, r_3) & = 20(r_2 + r_3) \\
        &- \frac{80}{\pi}\int_{r_3}^1\sqrt{\frac{(1-\mu)^3}{\mu(\mu-r_2)(\mu-r_3)}}d\mu,
    \end{split}
\end{equation}
so, according to Eq.~\eqref{eq31}, the Riemann invariants $r_2$ and $r_3$ must satisfy the system
\begin{equation} \label{eq73}
    w_2^B(r_2, r_3) - w_3^B(r_2, r_3) + (v_3(r_2, r_3) - v_2(r_2, r_3))t = 0,
\end{equation}
where the values of $w^B_2(r_2, r_3)$ and $w^B_3(r_2, r_3)$ can be calculated from Eq.~\eqref{eq33} with $W^B(r_2, r_3)$ defined by Eq.~\eqref{eq72}. We solved Eq.~\eqref{eq73} numerically by finding $r_2$ for any given $r_3$. The phase correction is given by (see Eq.~(\ref{eq55}))
\begin{equation} \label{eq74}
    \theta_0(r_2, r_3) = - k W^B(r_2, r_3),
\end{equation}
where $k = 2\sqrt{r_3}$ and $C=0$ since the initial phase shift is equal to zero.

\begin{figure}[t]
    \centering
    \includegraphics[width=7cm]{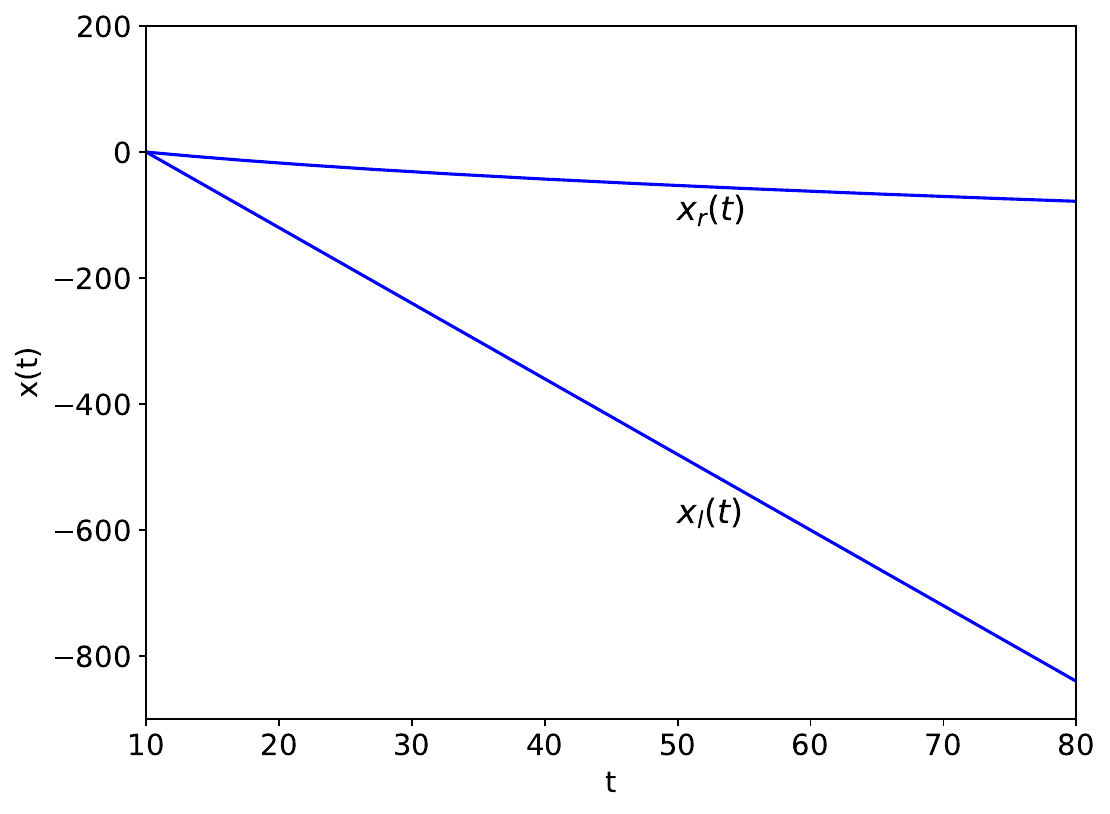}
    \caption{Trajectories of the soliton edge $x_r(t)$ and the small-amplitude edge $x_l(t)$, determined by Eqs.~\eqref{eq76} and \eqref{eq77}, respectively.}
    \label{fig4}
\end{figure}

\subsection{Evolution of the DSW and numerical validation}

\begin{figure}[t]
    \centering
    \includegraphics[width=6cm]{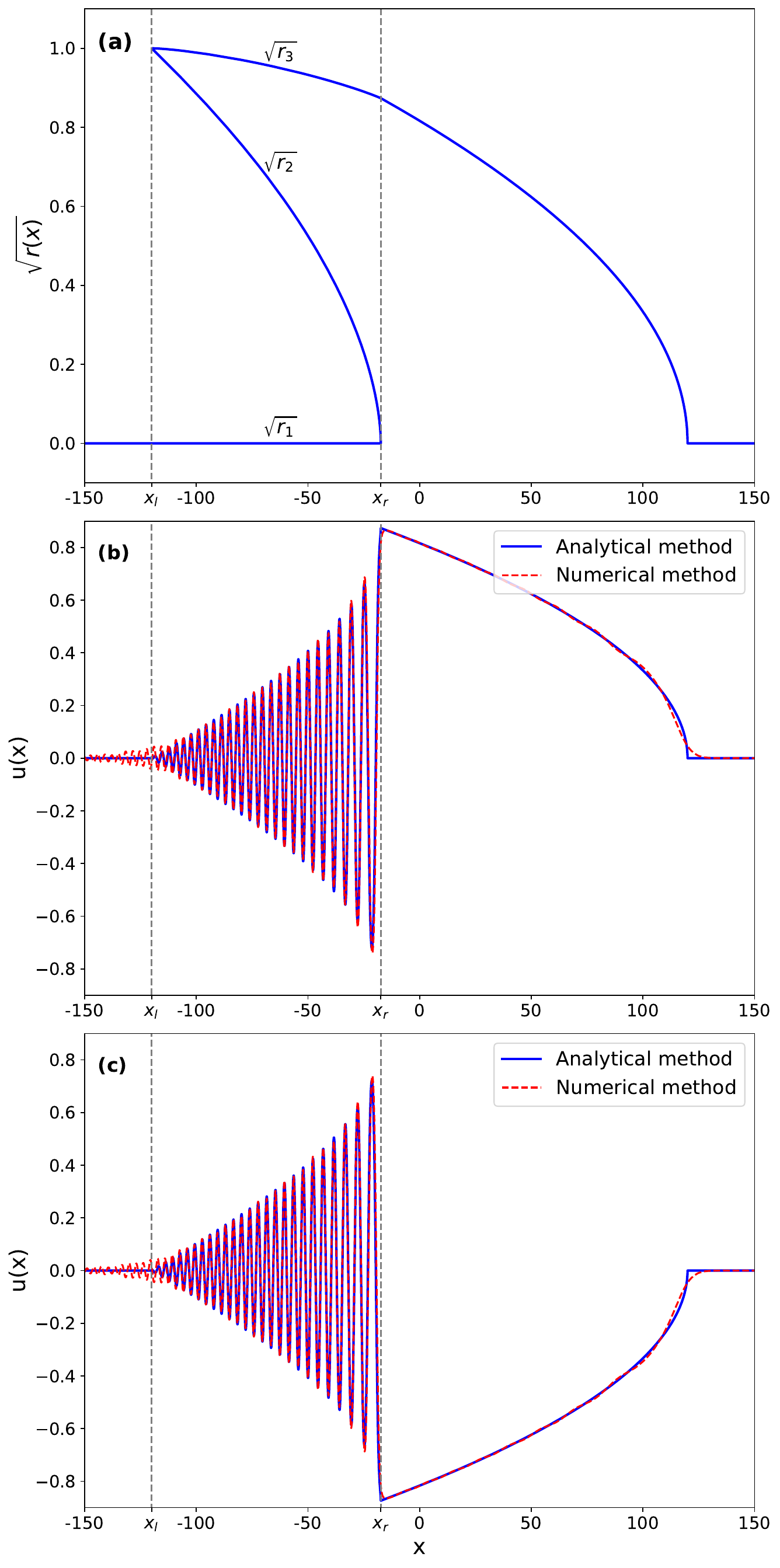}
    \caption{Analytical and numerical solutions of the mKdV equation evolved from the initial distribution \eqref{eq57} at $t = 20$. (a) Profiles of the Riemann invariants used to analytically construct the DSW distribution via Eq.~\eqref{eq6}. The resulting wave profiles are shown for (b) positive and (c) negative initial pulses, corresponding to the two signs in Eq.~\eqref{eq57}. Analytical results (solid blue lines) are compared with numerical solutions obtained via the split-step Fourier method (dashed red lines).}
    \label{fig5}
\end{figure}

The comparison of analytical and numerical solutions for a positive initial distribution is shown in Fig.~\ref{fig3}(d)-(f). In this case, the soliton edge moves along the second branch of the dispersionless solution of the Hopf equation (see Eqs.~(\ref{eq2}) and (\ref{eq3})). Moreover, at the instant of breaking, the soliton edge moves through the background $r_0 = 1$; therefore, solving Eq.~\eqref{eq23}, we find
\begin{equation} \label{eq75}
    r_r(t) = \left( \frac{20}{10 + t} \right)^{\frac{2}{3}}.
\end{equation}
Substituting this formula into Eq.~\eqref{eq24} yields the path of the soliton edge
\begin{equation} \label{eq76}
    x_r(t) = 120 - 6\cdot20^{2/3}\left( t + 10 \right)^{1/3}.
\end{equation}
Eq.~\eqref{eq29} gives at once the path of the small-amplitude edge
\begin{equation} \label{eq77}
    x_l(t) = -12(t - 10).
\end{equation}
The plots of these two functions are shown in Fig.~\ref{fig4}, and they agree very well with the numerical results.
Within the DSW region, the distributions of the Riemann invariants are given by Eqs.~(\ref{eq31}), (\ref{eq33}), \eqref{eq73}, and their plots are shown in Fig.~\ref{fig5}(a).
As was mentioned earlier, the same Riemann invariants determine the distributions of the physical variable $u$ (calculated by Eq.~\eqref{eq6}) for both positive and negative initial symmetric pulses. These results are shown in Figs.~\ref{fig5}(b) and (c), respectively. As one can see, our analytical results agree perfectly well with the direct numerical simulations. 

Further results for a comparison of the analytical and numerical solutions for a DSW in both Regions $A$ and $B$ (see Fig.~\ref{fig2}(b)-(c)) can be found in Appendix~\ref{app:general_case}.

\section{Conclusion}

In this work, we have shown that the Gurevich-Pitaevskii approach based on Whitham modulation theory provides a very accurate description of the evolution of an initial pulse in the case of mKdV dynamics.
We considered situations where the initial pulses do not eventually evolve into trains of well-separated solitons, so our theory extends the results obtained earlier in Refs.~\cite{el93,el02,isokam19} for the
KdV case to the mKdV situation. We found that the Gurevich-Pitaevskii analytical approach agrees very well with the direct numerical simulations even within one wavelength scale after taking into account the phase shift correction. Thus, the Whitham modulation theory turns out to be very accurate almost everywhere along a DSW for a not-too-small time of evolution after the wave-breaking moment, so it must provide a quite reliable description of DSWs observed in real experiments. This theory can find applications to observations of dispersive shocks in magneto-hydrodynamic nonlinear Alfv\'{e}n waves and in two-component Bose-Einstein condensates, where the wave dynamics can be reduced to the mKdV equation (see, e.g., \cite{kamchbook,mjolhus89,kklp14}).


\begin{acknowledgments}

L. F. Calazans de Brito acknowledges financial support from the Fundação de Amparo à Pesquisa do Estado
de São Paulo (FAPESP), Grant No. 2023/17459-8. A. Gammal acknowledges support from FAPESP,
Grant No. 2024/01533-7, and from the Conselho Nacional de Desenvolvimento Científico e Tecnológico (CNPq),
Grant No. 306219/2022-0.

\end{acknowledgments}

\appendix
 
\section{Evolution of a general quasi-simple wave pulse}
\label{app:general_case}

\begin{figure*}[t]
	\centering
	\includegraphics[width=17cm]{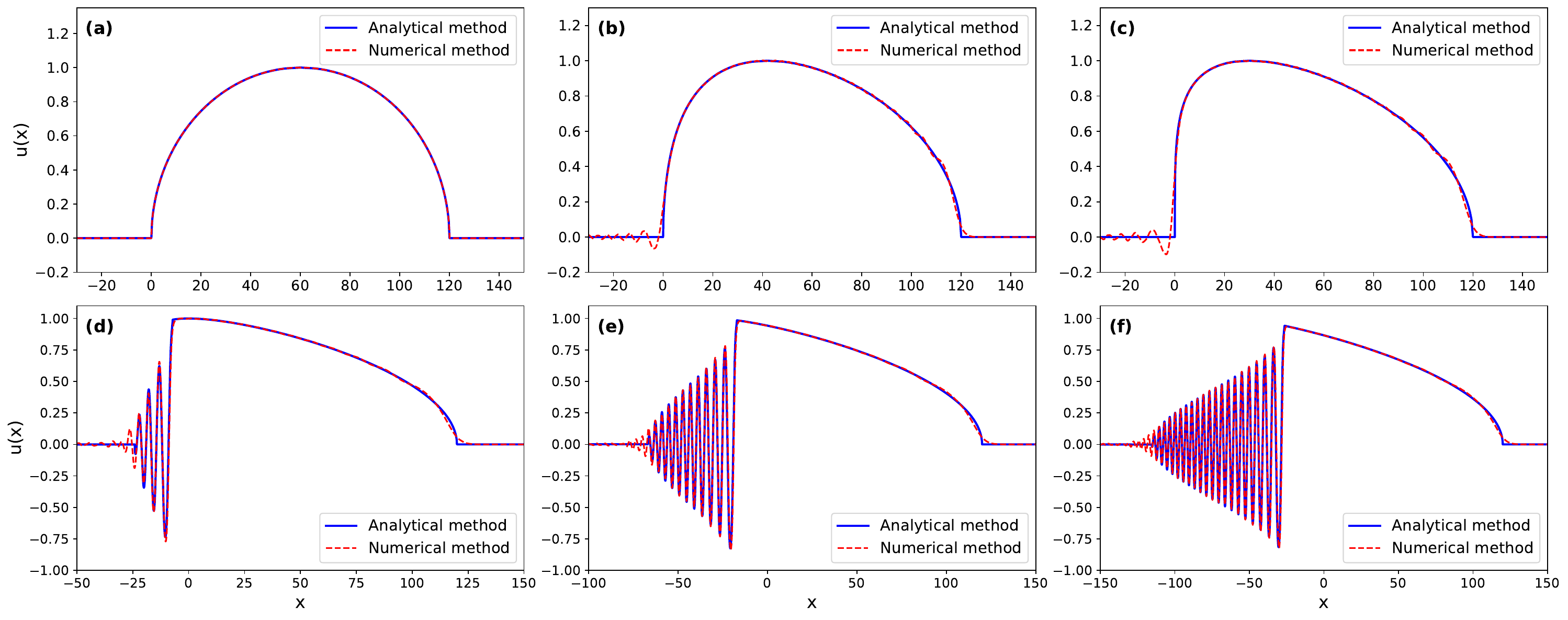}
	\caption{Evolution of a general quasi-simple positive pulse along regions $A$ and $B$ for the mKdV equation at times (a) $t=0$, (b) $t=3$, (c) $t=5$, (d) $t=10$, (e) $t=15$, and (f) $t=20$. The plots show a comparison between numerical results obtained via the split-step Fourier method (dashed red lines) and analytical solutions (solid blue lines) derived from Eq.~\eqref{eq6} using the Gurevich-Pitaevskii theory.}
	\label{fig6}
\end{figure*}

In this section, we discuss in detail the analytical theory of a general wave-breaking situation that evolves into dispersive shock waves in both Regions $A$ and $B$. We compare our analytical predictions with numerical solutions, which are obtained using the same methodology and parameters described in Section~\ref{comparison}.

An example of such a pulse that evolves into both Regions $A$ and $B$ (see Fig.~\ref{fig6}) is defined by the initial distribution
\begin{equation}\label{eqA1}
	u(x, 0) =
	\begin{cases}
		\pm \frac{1}{60}\sqrt{x(120 - x)}, \qquad 0 \leq x \leq 120 \\
		0, \qquad \text{otherwise},
	\end{cases}
\end{equation}
which is simple enough for explicit calculations.

\subsection{Non-dispersive regime}

Before the wave-breaking time, in the non-dispersive regime~(see Fig.~\ref{fig6}(a)-(c)), the initial conditions for the two branches can be found by inverting Eq.~\eqref{eqA1}
\begin{equation}
	\bar{x}_1(r) = 60(1 - \sqrt{1 - r}), \quad \bar{x}_2(r) = 60(1 + \sqrt{1 - r}),
\end{equation}
where we define $r(x,t) = u^2(x,t)$. In this stage, the pulse evolution satisfies the Hopf equation~\eqref{eq3}, yielding two families of characteristics
\begin{equation}
	\begin{split}
		x_1(r,t) = 60(1 - \sqrt{1 - r}) - 6rt,  \\
		x_2(r,t) = 60(1 + \sqrt{1 - r}) - 6rt.
	\end{split}
\end{equation}
This solution loses its validity at the wave-breaking time. Using Eq.~\eqref{eq4} and noting that $\text{max}(dr/dx) = 1/30$, this critical time is evaluated as
\begin{equation}
	t_{\text{WB}} = 5.
\end{equation}
Following wave breaking, the DSW emerges and begins its expansion.

\subsection{DSW parameters in Region $A$}
\begin{figure}[t]
	\centering
	\includegraphics[width=6cm]{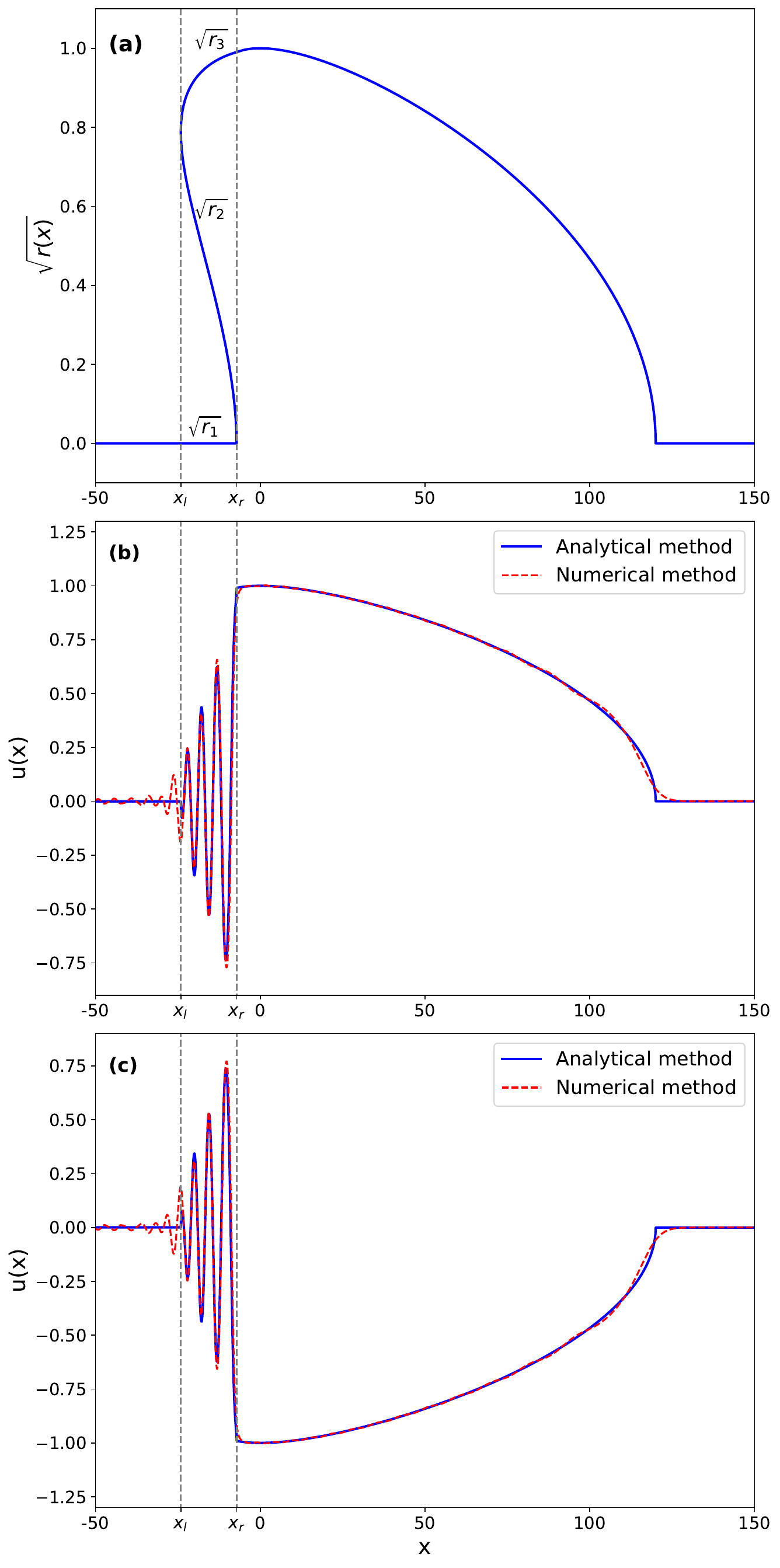}
	\caption{Analytical and numerical solutions of the mKdV equation evolved from the initial distribution \eqref{eqA1} at $t = 10$. (a) Profiles of the Riemann invariants in region $A$, used to analytically construct the DSW distribution via Eq.~\eqref{eq6}. The resulting wave profiles are shown for (b) positive and (c) negative initial pulses, corresponding to the two signs in Eq.~\eqref{eqA1}. Analytical results (solid blue lines) are compared with numerical solutions obtained via the split-step Fourier method (dashed red lines).}
	\label{fig7}
\end{figure}

After the wave-breaking time, and before the soliton edge achieves the background maximum value at $r_r = 1$, the entire DSW structure is contained within Region $A$ (see Fig.~\ref{fig7}). By evaluating the parameter $\Psi^A(\mu)$ from Eq.~\eqref{eq46},
\begin{equation} \label{eqA5}
	\Psi^A(\mu) = \frac{30}{\pi}\left[1 - \frac{1 - \mu}{\sqrt{\mu}}\arctan{(\sqrt{\mu})} \right],
\end{equation}
we find that the potential function defined in Eq.~\eqref{eq47} takes the form
\begin{equation} \label{eqA6}
W^A(r_2, r_3) = 30\left[1 - \frac{1}{\pi}\int_{r_2}^{r_3}\frac{(1 - \mu) \arctan{(\sqrt{\mu})}}{\sqrt{\mu(r_3 - \mu)(\mu - r_2)}}d\mu \right].
\end{equation}

\subsubsection{Soliton edge parameters in Region $A$}

We can determine the parameters of the soliton edge by solving Eq.~\eqref{eq23}. At the wave-breaking time~($t = 5$), the shock wave edges begin to propagate from $r_0 = 0$. Consequently, solving Eq.~\eqref{eq23} yields
\begin{equation}\label{eqA8}
	t(r_r) = \frac{15}{2r_r}\left(\frac{\arcsin{(\sqrt{r_r})}}{\sqrt{r_r}}  - \sqrt{1 - r_r}\right),
\end{equation}
and while the soliton edge propagates within Region A, its position is given by Eq.~\eqref{eq24}
\begin{equation}
	x_r(r_r) = -6r_rt(r_r) + \bar{x}_1(r_r).
\end{equation}
As previously mentioned, the soliton edge enters Region $B$ when it reaches the background maximum at $r_r = 1$. The transition time $t_{\text{AB}}$ can thus be evaluated using Eq.~\eqref{eqA8}
\begin{equation} \label{eqA10}
	t_{\text{AB}} = \frac{15}{4} \pi.
\end{equation}
The analytical description of the soliton edge dynamics in Region $B$ will be considered in more detail in the following subsection.

\subsubsection{Linear limit edge parameters}

In contrast to the problem addressed in the main text, where the linear limit edge propagates at a constant group velocity, a more general situation involves a time-dependent velocity, requiring the generalized hodograph method to determine this edge dynamics. At this limit, $r_3 = r_2 = a$, and both equations of the system \eqref{eq31} reduce to
\begin{equation}\label{eqA11}
	x + 12at = w^A(a,a).
\end{equation}
Applying the relation
\begin{equation}
\begin{split}
	\frac{\prt W^A (r_2, r_3)}{\prt r_3}\bigg|_{r_2 = r_3 = a}& = \frac{\prt W^A (r_2, r_3)}{\prt r_2}\bigg|_{r_2 = r_3 = a}\\
 &= \frac{1}{2}\frac{d W^A (a, a)}{d a}
\end{split}
\end{equation}
to Eq.~\eqref{eq33}, we find that
\begin{equation} \label{eqA13}
	w^A(a,a) = \left( 1 + 2a \frac{\partial}{\partial a} \right) W^A(a,a).
\end{equation}
The potential $W^A(a,a)$ can be evaluated as the limit
\begin{equation}
	W^A(a,a) = \lim_{\eps \rightarrow 0} \int_{a}^{a+\eps} \frac{\Psi^A(\mu) d \mu}{\sqrt{(a + \eps -\mu)(\mu - a)}} \approx \pi \Psi^A(\mu).
\end{equation}

Using Eq.~\eqref{eqA5}, we obtain
\begin{equation} \label{eqA15}
	W^A(a,a) = 30\left[1 - \frac{1 - a}{\sqrt{a}}\,\arctan{(\sqrt{a})} \right],
\end{equation}
and consequently, substituting Eq.~\eqref{eqA15} into Eq.~\eqref{eqA13} yields
\begin{equation}
	w^A(a,a) = 60 \sqrt{a}\, \text{arctanh}(\sqrt{a}).
\end{equation}

The time, at which the linear limit reaches the point corresponding to the parameter $a$, can be found by differentiating Eq.~\eqref{eqA11} with respect to $t$, so we get the expression

\begin{equation} \label{eqA17}
	t(a) = \frac{1}{12} \frac{d\omega}{da} = \frac{5}{2} \left[ \frac{1}{\sqrt{a}} \text{arctanh}(\sqrt{a}) + \frac{1}{1-a} \right].
\end{equation}
As can be seen from Eq.~\eqref{eqA17}, since $t(a = 1) \rightarrow \infty$, the linear limit edge never gets into the region $B$. So, its position $x_l(r_l, t)$ can always be found from Eq.~\eqref{eqA11}.

\subsubsection{Riemann invariants distribution in Region $A$} \label{riemannRegionA}

Knowing the position and parameters of the soliton and linear limit edges, we can calculate the values of the Riemann invariants from the system~\eqref{eq31}. Consequently, $r_2$ and $r_3$ must satisfy the relation
\begin{equation}\label{eqA7}
	w_2^A(r_2, r_3) - w_3^A(r_2, r_3) + (v_3(r_2, r_3) - v_2(r_2, r_3))t = 0.
\end{equation}
The functions $w_2^A(r_2, r_3)$ and $w_3^A(r_2, r_3)$ are calculated by substituting Eq.~\eqref{eqA6} into Eq.~\eqref{eq33}. In this regime, $r_3 \in (r_l, r_r)$, and the coordinate of the soliton edge is smaller than the coordinate of the maximum background value with $r=1$ (i.e., $x_r < \bar{x}_1(1) = 60 - 6t$). Since the domain of $r_3(x)$ is known, we solve numerically Eq.~\eqref{eqA7} to find the corresponding value of $r_2(x)$. The distribution of the Riemann invariants for a DSW in Region $A$ is shown in Fig.~\ref{fig7}(a), and the resulting wave profiles for both positive and negative initial conditions are again described by Eq.~\eqref{eq6}, as illustrated in Fig.~\ref{fig7}(b)-(c). The phase correction for the region $A$ is given by Eq.~(\ref{eq55})
\begin{equation} \label{eqAPC}
	\theta_0(r_2, r_3) = - k W^A(r_2, r_3),
\end{equation}
where $k = 2\sqrt{r_3}$.

\subsection{DSW parameters in Region $B$}

\begin{figure}[t]
	\centering
	\includegraphics[width=6cm]{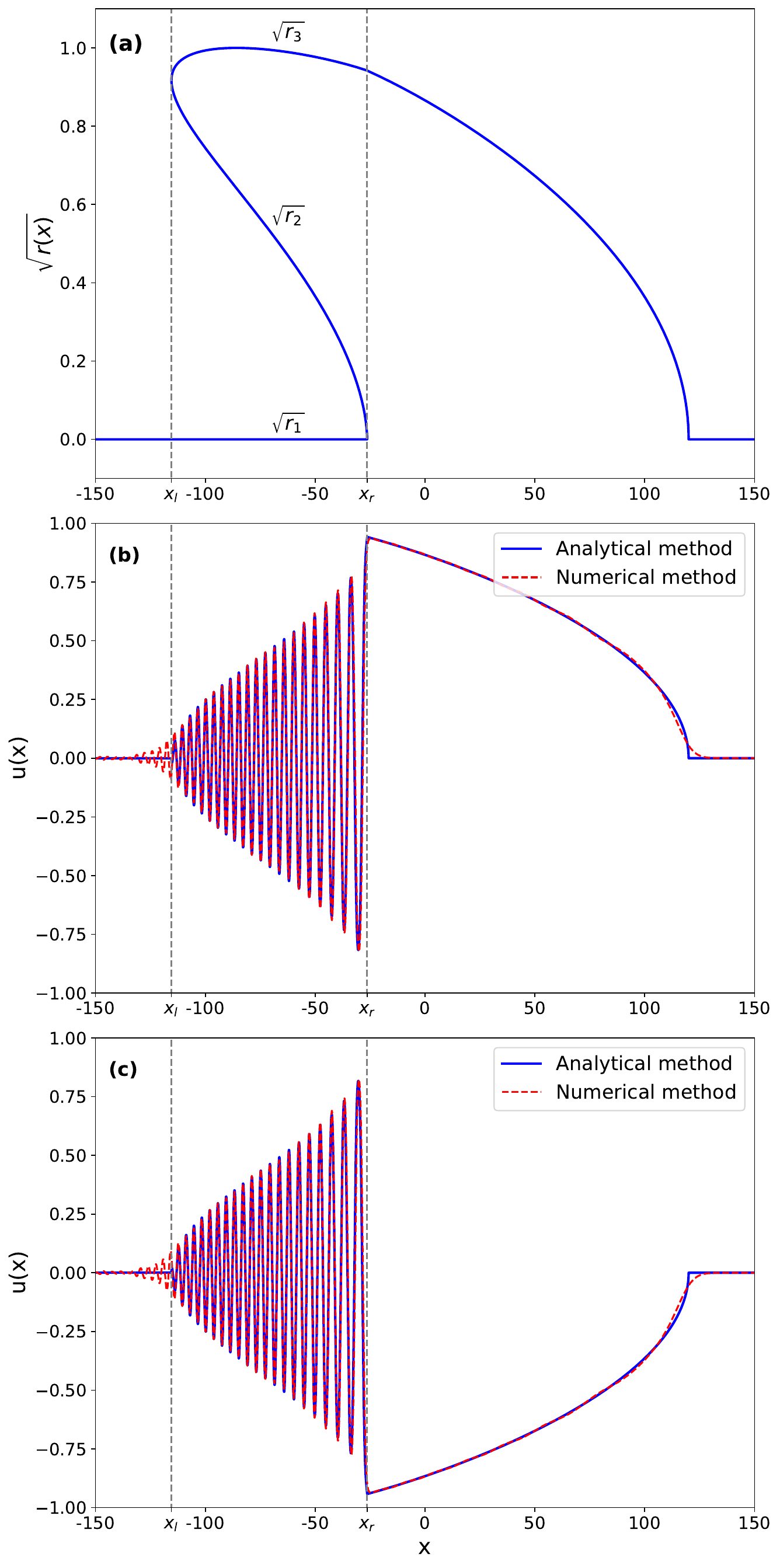}
	\caption{Analytical and numerical solutions of the mKdV equation for the DSW evolved from the initial distribution \eqref{eqA1} at $t = 20$. (a) Profiles of the Riemann invariants in Regions $A$ and $B$ are used to analytically construct the DSW distribution via Eq.~\eqref{eq6}. The resulting wave profiles are shown for (b) positive and (c) negative initial pulses, corresponding to the two signs in Eq.~\eqref{eqA1}. Analytical results (solid blue lines) are compared with numerical solutions obtained via the split-step Fourier method (dashed red lines).}
	\label{fig8}
\end{figure}

When the soliton edge enters the Region B~(see Fig.~\ref{fig8}), the pulse parameters are calculated differently in the Regions $A$ and $B$. In particular, in Region $A$ we have the condition $r_3 \in (a, 1)$, while in Region $B$ we have $r_3 \in (1, r_r)$. From Eq.~\eqref{eq50}, we find that
\begin{equation}
	\Psi^B(\mu) = - \frac{ 30(1 - \mu)}{\sqrt{\mu}}.
\end{equation}
Then, the potential function in Region $B$ can be calculated using Eq.~\eqref{eq48}
\begin{equation}
	W^B(r_2, r_3) = W^A(r_2, r_3) - 30\int_{r_3}^{1}\frac{(1-\mu )d\mu}{\sqrt{\mu(\mu - r_2)(\mu - r_3)}}.
\end{equation}

\subsubsection{Soliton edge parameters in Region $B$}

The dynamics of the soliton edge in Region $B$ is governed by the differential equation~\eqref{eq22}. However, instead of using Eq.~\eqref{eq23}, which was derived for a soliton limit propagating entirely along a single branch for $t > t_{\text{WB}}$, the transition to the branch $\bar{x}_2(r)$ requires as new initial conditions $t_0 = t_{\text{AB}}$~(Eq.~\eqref{eqA10}) and $r_0 = 1$. This yields the integral expression
\begin{equation}
	t(r_r) = \left(\frac{1}{r_r}\right)^{3/2} t_{\text{AB}} + \frac{1}{4r_r^{3/2}}\int_{1}^{r_r} \sqrt{r}\frac{d\bar{x}_2(r)}{dr}dr.
\end{equation}
Consequently, the explicit trajectory for the soliton edge in Region $B$ takes the form
\begin{equation}\label{eqA21}
	t(r_r) = \frac{15}{2r_r} \left[ \frac{1}{\sqrt{r_r}}\left( \pi - \arcsin{(\sqrt{r_r})} \right) + \sqrt{1 - r_r} \right],
\end{equation}
and its position is found from
\begin{equation}
	x_r(r_r) = -6r_rt(r_r) + \bar{x}_2(r_r).
\end{equation}

\subsubsection{Riemann invariants distribution after the soliton edge reaches Region $B$}

As mentioned before, when the soliton edge enters Region $B$ (i.e., for $t > t_{\text{AB}}$), the distribution of the Riemann invariants can be determined across the entire DSW taking into account both regions. In Region $A$ ($r_3 \in (a, 1)$), the values of $r_2(x)$ are still calculated by Eq.~\eqref{eqA7}, as shown in Eq.~\ref{riemannRegionA}, and the phase correction satisfies Eq.~\eqref{eqAPC}. Meanwhile, in Region $B$ ($r_3 \in (1, r_r)$), the function $r_2(x)$ is found numerically from the relation
\begin{equation}\label{eqA23}
	w_2^B(r_2, r_3) - w_3^B(r_2, r_3) + (v_3(r_2, r_3) - v_2(r_2, r_3))t = 0,
\end{equation}
and the phase correction is given by
\begin{equation}
	\theta_0(r_2, r_3) = - k W^B(r_2, r_3).
\end{equation}

The distributions of the Riemann invariants for the pulse with initial condition~\eqref{eqA1} at $t=20$ are shown in Fig~\ref{fig8}(a), and the wave distribution for both positive and negative initial profiles can be determined from Eq.~\eqref{eq6}~(see Figs.~\ref{fig8}(b)-(c)). As one can see, the approximate analytical theory agrees very well with the numerical solutions.

\bibliographystyle{apsrev4-2}
\bibliography{bib.bib}

\end{document}